\documentclass[a4paper]{jpconf}
\usepackage{graphicx}
\usepackage[utf8]{inputenc}

\begin{document}
\title{Physics with charmonia at the SPD experiment}

\author{Igor Denisenko (on behalf of the SPD working group)}

\address{Joint Institute for Nuclear Research, 6 Joliot-Curie, 141980, Dubna, 
	Moscow region, Russia}

\ead{iden@jinr.ru}

\begin{abstract}
The $J/\psi$ production is a powerful probe of hadron structure, which is 
complementary to other parts of the SPD physics program. In this work, 
the current experimental status and modern theoretical approaches  to 
$J/\psi$ and charmonia production are reviewed. In this context,
the SPD performance and feasibility of key measurements are discussed.
\end{abstract}

\section{Introduction}

The spin-dependent parton structure is a fundamental property of a proton, which
might be a key to resolve famous ``spin crisis''.
 It has been a subject of
dedicated studies for several decades but our understanding remains
fragmentary and the worst known piece of the puzzle are spin-dependent
parton distributions of gluons.
The current situation is expected to be significantly improved
by a new experiment with polarized beams at the NICA accelerator complex. 
The proton-proton and proton-nucleon collisions will be collected by
the Spin Physics Detector (SPD)~\cite{Savin:2014sva}. The CMS energy for
proton-proton collisions at NICA varies from 10~GeV to 27~GeV.

There are several physical tools to probe the internal structure of the nucleon:
the Drell-Yan process, prompt photons, charmonia production, and inclusive hadron
production. Here the physics with charmonia is considered.
On the parton level, the $J/\psi$ production is expected to be dominated by gluon-gluon
fusion but the contribution of quark-antiquark annihilation may be significant
in some kinematic regions. This makes inclusive charmonia production complementary
to prompt photons and Drell-Yan studies.
From the experimental point of view, the simplest charmonia state to study is
$J/\psi$ due to a large
production cross-section and a clean experimental signature in the dimuon decay mode.
It has the same observables as the Drell-Yan process. For the unpolarized case, they
are the total cross-section, the $J/\psi$ $p_T$ and $x_F$ spectra and polarization.
For polarized hadron collisions, the transverse structure of hadrons can be
probed by measuring spin asymmetries of the $J/\psi$ production or studying the cross-section angular modulations. In particular, the transverse single
spin asymmetries (TSSA) are associated with the Sivers effect for gluons (e.g.
see~\cite{DAlesio:2017rzj}). Probing quark transverse momentum dependent parton distribution functions
through $J/\psi$ spin asymmetries might be also possible (similarly to what
is suggested in Ref.~\cite{Anselmino:2004ki} for the $p\bar{p}$ collisions)
but depends on the relative contribution
of the quark-antiquark annihilation process.
The challenging part is the interpretation of experimental
results due to the lack of theoretical understanding of the charmonia production
process. The interpretation is further complicated due to the fact that about 40\%
$J/\psi$ mesons are produced not directly in parton-parton interaction but through
so-called ``feed-down'' decays of $\chi_{c1,c2}\to\gamma J/\psi$ and $\psi(2S)\to
\pi\pi J/\psi $. 

Compared to the previous mostly fixed-target experiments, SPD
is expected to have several advantages. It is designed as
an open spectrometer with a high momentum resolution and allows studying $J/\psi$ alongside with $\chi_{c1}$, $\chi_{c2}$ and $\psi(2S)$.
Moreover, the high expected statistics should allow for a precise measurement
of $\psi(2S)$ production properties, that does not suffer from
the feed-down contributions, as well as provide an improved measurement
of $J/\psi$ TSSA.  In general, the experiment is expected
to validate current theoretical approaches and use them to interpret
observables for polarized beams. As a by-product, the validated
theoretical models can be used to extract pion and kaon gluon parton
distribution functions in future experiments like AMBER~\cite{Denisov:2018unj}.

\section{Available experimental results}
\begin{figure}[h]
	\begin{center}
		\begin{minipage}{19pc}
			\includegraphics[width=\textwidth]{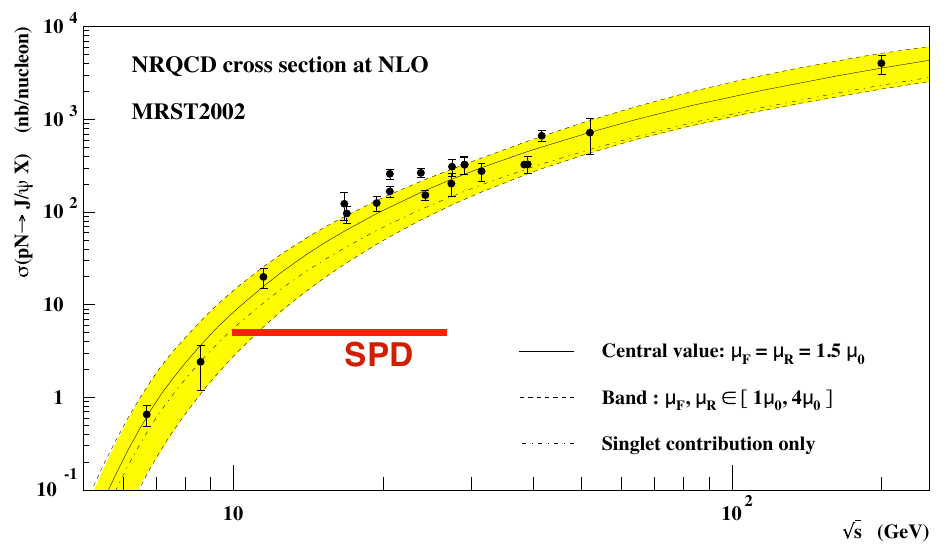}
			\caption{\label{fig:jpsi_cs}Inclusive $J/\psi$ production cross-section as a function of $\sqrt{s}$ in proton-proton and proton-nucleon collisions from Ref.~\cite{Maltoni:2006yp}.
			The SPD energy range is shown by the red line.
			}
		\end{minipage}
		\hspace{2pc}%
		\begin{minipage}{16pc}
			\begin{center}
				\includegraphics[width=.7\textwidth]{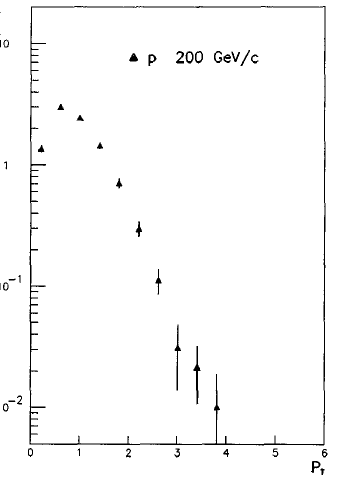}
			\end{center}
   		\caption{\label{fig:jpsi-pt} The $p_T$ spectra of $J/\psi$ measured by NA3
   		at $\sqrt{s}= 19.4$~GeV~\cite{Badier:1983dg}.}
		\end{minipage}
	\end{center}
\end{figure}

%\subsection{Inclusive $J/\psi$ production}
A good review of experimental results on the inclusive $J/\psi$ production
cross-section in
proton-proton and proton-nucleon collisions at low energies can be found
in Ref.~\cite{Maltoni:2006yp}
The cross-section measurements corrected for the nuclear dependence
and the SPD energy range are shown in Fig.~\ref{fig:jpsi_cs}.
The results are mostly obtained in fixed-target experiments
with a pion absorber, thus making it impossible to study $\chi_{cJ}$
and conduct a detailed investigation of $\psi(2S)$ production properties.
It must also be noted that shown results are not well-consistent.
For the highest available energies at SPD, the cross-section is about
200-250~nb.

The $p_T$ spectrum obtained by the NA3 collaboration at $\sqrt{s}=19.4$
in $pp$ collisions is shown in Fig.~\ref{fig:jpsi_pt}. It shows that typical transverse momenta are
of order of 1~GeV/$c$ and is comparable to $M_{J/\psi}c$. The $x_F$ distribution
at low energies is measured by NA3~\cite{Badier:1983dg}, E705~\cite{Antoniazzi:1992af}, and E866~\cite{Vogt:1999dw} to validate
relative contributions from gluon-gluon fusion and quark-antiquark annihilation
in the model (or to phenomenologically separate them). The worst known
observable, and the most interesting for the model validation, is
$J/\psi$ polarization. At these energies, its kinematic dependence
on $p_T$ and $x_F$ is measured by two experiments E866~\cite{Chang:2003rz}
and HERA-B~\cite{Abt:2009nu}.	
The results are notably different but cover different kinematic
regions.

For the polarized beams, the PHENIX collaboration reported nonzero TSSA
in the forward $x_F$ region with statistical significance of 3.3$\sigma$
based on data set of approximately 22,000 collected $J/\psi$ events~\cite{Adare:2010bd}.
The indication of nonzero TSSA are also reported in a recent publication
of this collaboration~\cite{Aidala:2018gmp}.

%\subsection{$\chi_{c1}$ and $\chi_{c2}$ production}
Decays of $\chi_{c1}$ and $\chi_{c2}$ are known to contribute about
30\% of inclusive $J/\psi$ events. These states are poorly known:
their production cross-sections are measured with huge uncertainties
and their polarization has not been measured yet. Their relative contribution
to $J/\psi$ decays is especially essential for validation of
theoretical models but also remains quite uncertain. The known experimental
results on the states at low energies are summarized in Ref.~\cite{Abt:2008ed}.

%\subsection{$\psi(2S)$ production}
Production properties of $\psi(2S)$ at the SPD energies are unknown except
for the total cross-section due to limitations of previous fixed-target
experiments. Proton-nucleon experimental results are reviewed
in Ref.~\cite{Maltoni:2006yp} the production cross-section is about 15\%
of the one for $J/\psi$.

\section{Charmonia production models}
There are two contemporary models for inclusive $J/\psi$ production:
Color Evaporation Model (CEM) and
Non-Relativistic QCD (NRQCD). In both models, charmonia state hadronizes from
a perturbatively produced $c\bar{c}$ pair, but there is a notable difference
in the hadronization process, number of free non-perturbative parameters and
predicted observables. Both models are mostly used with the collinear
factorization, which is assumed in the following review of models.
As it will be discussed later, this factorization is not
appropriate for SPD energies, so possible alternatives will be discussed later.

\begin{figure}[h]
	\begin{center}
		\includegraphics[width=.8\textwidth]{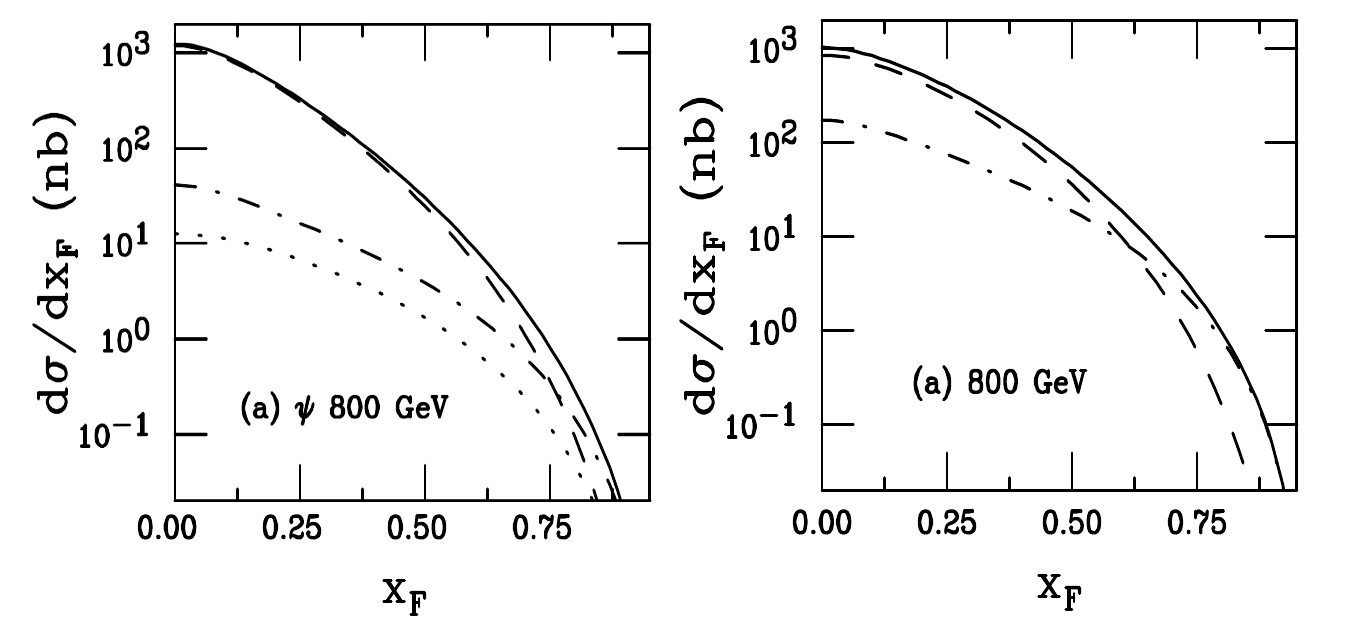}
		\caption{\label{fig:nrqcd-cem}NRQCD (left) and CEM (right) predictions of $x_F$ spectra in 
			$pp$ collisions at $\sqrt{s}=39$~GeV from Ref.~\cite{Vogt:1999dw}. The dashed and dashed-and-doted
			lines show gluon-gluon fusion and quark-antiquark annihilation, respectively.}
	\end{center}
\end{figure}

Before proceeding to description of the models, the difference in their predictions
for the relative contribution the gluon-gluon fusion and quark-antiquark annihilation
in the $x_F$ spectrum is shown in Fig.~\ref{fig:nrqcd-cem}.

\subsection{Color evaporation model}
In CEM, production cross-section of a charmonia state is
assumed to be proportional to the one of a $c\bar c$ pair
with the invariant mass between $2m_c$ and the open charm threshold, where
$m_c$ is the mass of $c$-quark.
In this model, the sum over colors (they are assumed to be neutralized
by emission of soft gluons) and spins of the quark and antiquark
is implied. Explicit formulas can be found e.g. in Ref.~\cite{Vogt:1999dw}.
The proportionality coefficients for a given charmonium
state are assumed to be process-independent. 
Despite its simplicity, this phenomenological model provides
reasonably good descriptions of $\sqrt{s}$-dependence, $d\sigma/dx_F$
\cite{Gavai:1994in}. At NLO, it can also reasonably describe
$p_T$ distribution if $k_T$ smearing is considered~\cite{Bodwin:2005hm}.
At the same time, the process independence of
proportionality coefficients holds only approximately~\cite{Bodwin:2005hm}.
Even more importantly, the model predicts that all charmonia states
have the same shape of kinematic distributions, which is not supported
by experimental data~\cite{Adare:2011vq}.

\subsection{NRQCD}
The NRQCD~\cite{Bodwin:1994jh} is more rigorous approach to charmonia production.
It is based on two points, the first one is the factorization conjecture proven
only for sufficiently high $p_T$. It assumes that for a charmonia state
$H$ the inclusive production cross-section on the parton level can be written as
$$
\nonumber
%\label{eq:nrqcd}
\hat\sigma(ij\to H + X) = \sum_n C^{ij}_{n} \langle O^{H}_n \rangle.\nonumber
$$
Here indexes $i$ and $j$ denote interacting partons, $X$ states for
the rest of produced particles. The perturbative sort-distance
coefficients (SDC) $C^{ij}_{n}$ describe production of a $c\bar c$ pair
in the state $n$ ($n$ denotes their relative momentum, spin alignment and
color state) on the scale of $1/m_c$. Finally, the
non-pertubative long-distance matrix elements (LDME)
$\langle O^{H}_n \rangle$ describe hadronization of the produced $c\bar c$
pair to charmonia state $H$. An important consequence of this
factorization is the process independence for LDME (e.g. in hadron
collisions, photoproduction, $e^+e^-$ annihilation).
The second point is that there is a hierarchy of LDME $\langle O^H_n \rangle$
with respect to $v$ ($v^2\sim0.2-0.3$) -- typical velocity of heavy quark in
the charmonium  system. 
%Thus, Eq.~\ref{eq:nrqcd} is a double series in $\alpha_S$ and $v^2$.
%Both of these parameters are of order of 0.2-0.3, which may rise concerns
%about convergence of the series. 
%The color-singlet LDME are obtained
%in the same way as in CSM.
The color-octet LDME are obtained from fits
to the data. % or lattice calculations (the latter available only for 
%$\chi_{cJ}$ \cite{Bodwin:1996tg}).
The model predicts the full spectra
of observables including $J/\psi$ polarization.
Verification of the LDME universality in different $J/\psi$ production
processes is one of the direct tests of
NRQCD. The Ref.~\cite{Butenschoen:2012qr} reviews NLO NRQCD fits to different
data sets, that include the $p_T$ distribution in proton-proton collisions (ATLAS, CDF),
the same distribution in photoproduction at H1, the polarization data (CDF)
and the $J/\psi$ production in $e^+e^-$ annihilation at Belle showing that
all available data sets can not be described simultaneously by one set of LDME.
Inability to simultaneously describe $p_T$ spectra and polarization
may be related to the so-called "polarization puzzle" -- NRQCD difficulties
to describe the observed weak $J/\psi$ polarization. One of the possible
solutions to the puzzle is significant contribution of the feed-down decays.
It is based on the observed difference in polarization between $\Upsilon(1S)$
and its radial excitations $\Upsilon(2S)$ and $\Upsilon(3S)$
measured by E866~\cite{Faccioli:2012nv}.
%Their importance can be seen from the polarization measurements of $\Upsilon(1S)$
%and its radial excitations $\Upsilon(2S)$ and $\Upsilon(3S)$ performed
%by E866~\cite{Faccioli:2012nv}.
For charmonia Ref.~\cite{Faccioli:2018uik} suggests cancellation
of $\chi_{c1}$ and $\chi_{c2}$ contributions as a possible explanation,
while $\chi_{c1}$ production cross-section is usually strongly underpredicted
in NRQCD. New experimental data on $\chi_{cJ}$ contributions and their polarization,
as well as polarization measurements for $\psi(2S)$ are necessary to verify these ideas.
%Another problem of the reviewed fits is that they can not be extended to
%low $p_T$ values due to limitations of the collinear factorization.
%The typical low $p_T$ cut-off value in such NRQCD fits is 
%3--5~GeV/$c$ (or , which makes it inapplicable for SPD physics.
Another problem of the reviewed fits can be fitting 
of low-$p_T$ data, where collinear factorization may not
be applicable (typically $p_T$ spectra are fitted above 3~GeV or
larger cut-off values).

%At the same time, there may be other explanations of NRQCD difficulties to
%describe data. They include failure of fixed-order QCD expansion for SDC,
%slow convergence of NRQCD $v$-expansion or inclusion of small $p_T$
%data (below 3--5~GeV/$c$)
%to global fits with collinear factorization, where $k_T$ of partons become non-negligible.
%The typical low $p_T$ cut-off value in such NRQCD fits is around
%3--5~GeV/$c$, which makes it inapplicable for SPD physics.
% \cite{Brambilla:2014jmp}. %??

%The description of the  differential cross-section with respect to $x_F$
%is one the direct tests of charmomia prodcution at SPD energies.
%Previously, it was used by NA3 to separate the gluon-gluon fusion and
%quark-antiquark annihilation~\cite{Badier:1983dg} as well as by E537
%to probe gluon distributions of pion~\cite{Tzamarias:1990ij}.
%The Ref.~\cite{Vogt:1999dw} gives prediction of CEM and
%NRQCD models for collision energies of 15~GeV and 39~GeV. The
%total line shape and relative
%contributions of the gluon-gluon fusion and
%quark-antiquark annihilation are shown in Fig.~\ref{fig:nrqcd-cem}.
%

\subsection{Theoretical approaches for SPD}
The $p_T$ spectrum measured by NA3 at $\sqrt{s}=19.4$~GeV (see Fig.~\ref{fig:jpsi-pt})
shows that typical transverse momentum of $J/\psi$ at SPD is about 1~GeV/$c$
and its maximum expected value is about 4~-~5~GeV/$c$. The description of this
kinematics is challenging from the theoretical point of view and
requires taking into account transverse momentum of partons.
There are several approaches within the framework of NRQCD,
that may be suitable for SPD. Firstly, it is Parton Reggeization Approach~\cite{Kniehl:2006sk}, which is capable of describing the whole $J/\psi$ $p_T$ spectra
in high energy experiments  and has preliminary predictions for SPD.
Secondly, there is a set of works applying $k_T$-factorization to
charmonia production (see Ref.~\cite{Baranov:2015laa} and following
works of the authors).
It is important to outline that both these approaches have nontrivial
implications for other observables like $J/\psi$ polarization.
It has also been a recent attempt to significantly improve color
evaporation model~\cite{Ma:2016exq, Cheung:2018upe}. Despite good $p_T$ description, it fails to describe
$J/\psi$ polarization measured at HERA-B~\cite{Abt:2009nu}.

\section{Study of charmonia at SPD}
\label{sec:SPD}
\begin{figure}[h]
	\begin{center}
	\begin{minipage}{18pc}
	\begin{center}
		\includegraphics[width=\textwidth]{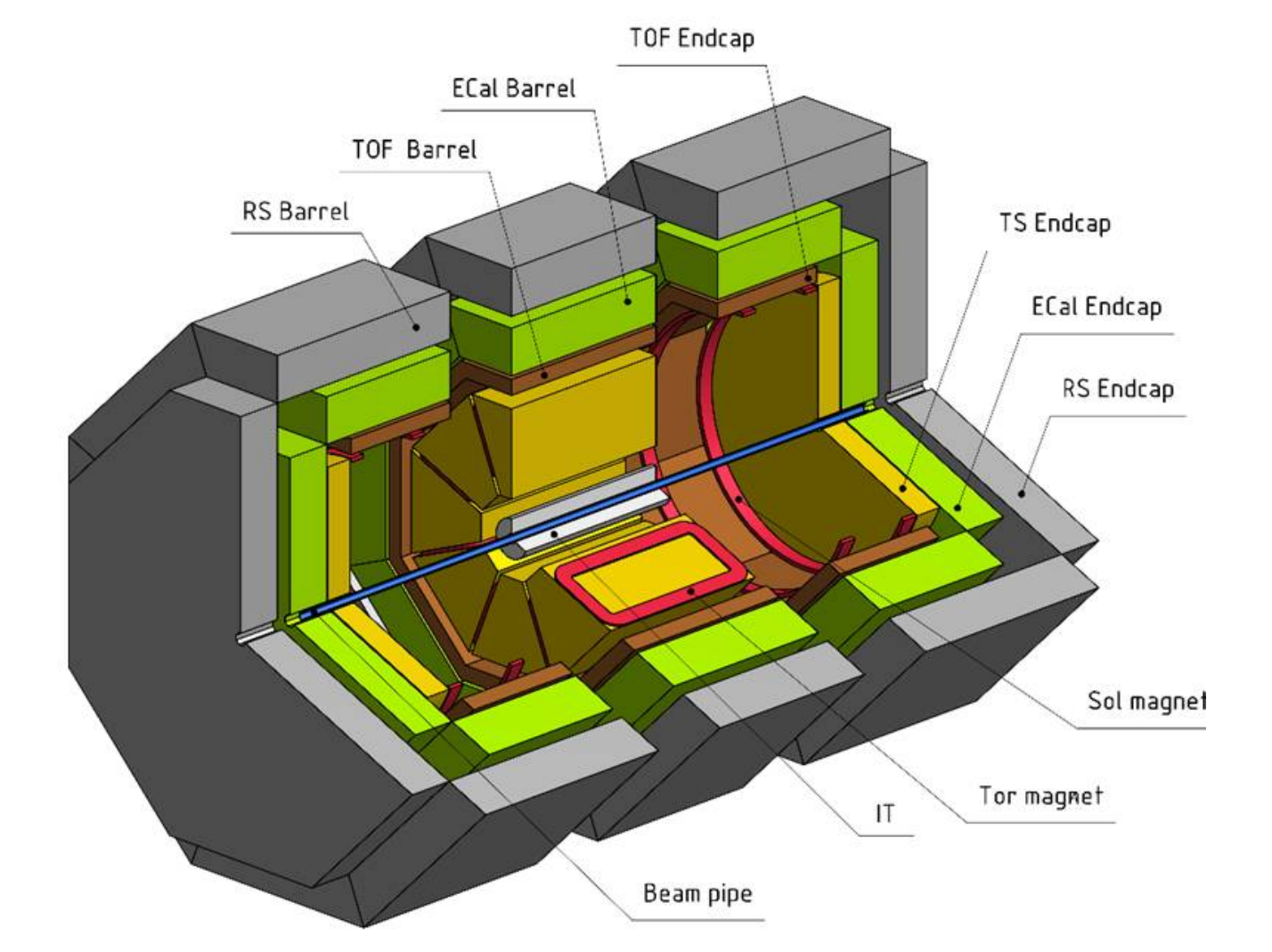}
			\caption{\label{fig:detector} Possible set-up of SPD~\cite{SPD_SDR}.}
	\end{center}
    \end{minipage} \hspace{1pc}
	\begin{minipage}{18pc}
	\begin{center}
		\includegraphics[width=\textwidth]{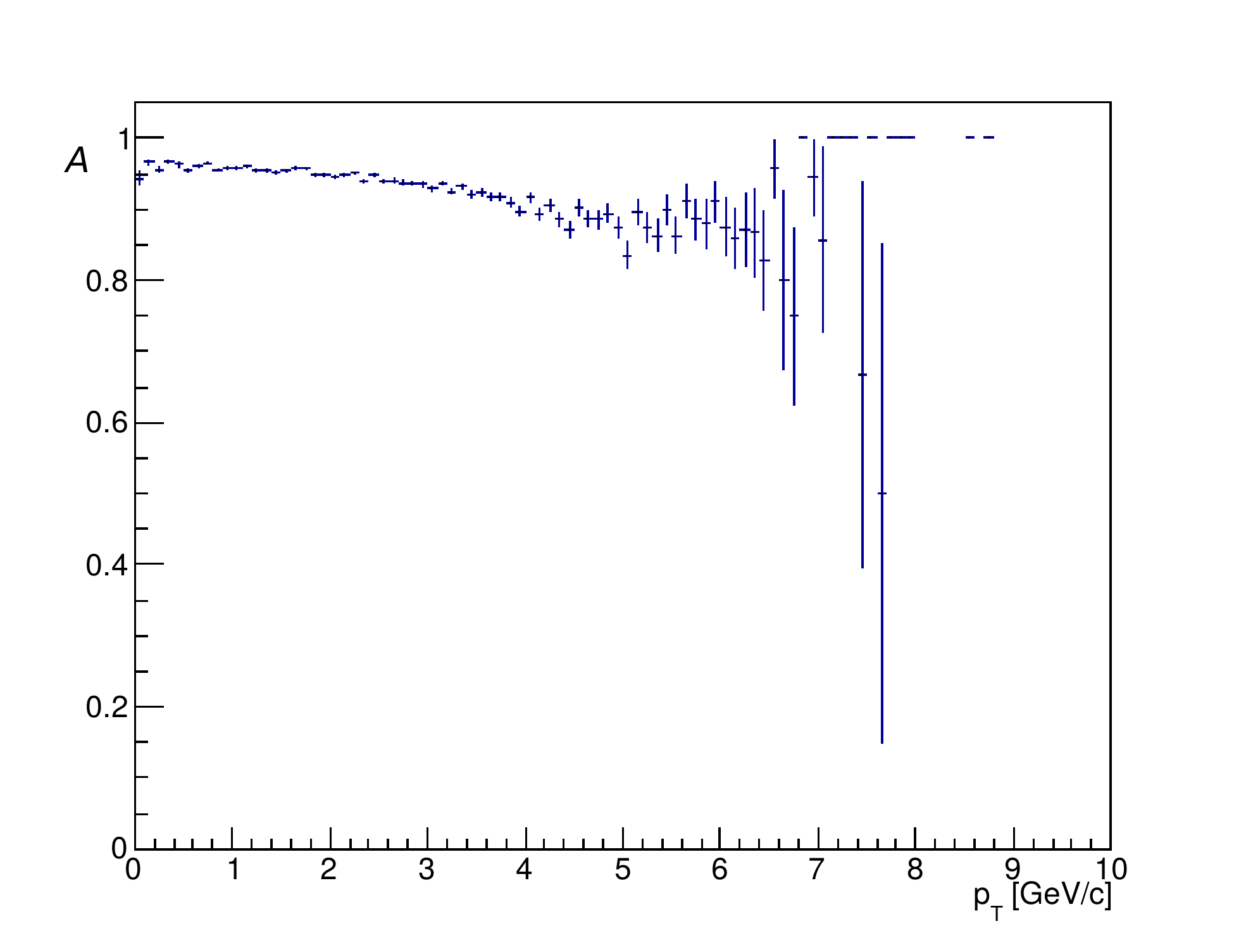}
		\caption{\label{fig:jpsi_pt} Detector acceptance for the $J/\psi$
		as a function of $p_T$.}
	\end{center}
    \end{minipage}
    \end{center}
\end{figure}
The final detector set-up has not been completed yet (see Ref.~\cite{SPD_SDR}
for the details and the designed performance). One of its possible configurations is
shown in Fig.~\ref{fig:detector}. The detector has close to $4\pi$ geometrical acceptance
and consists of a vertex detector, a straw tracker, an electromagnetic
calorimeter (ECAL) and a range system (RS) for muon identification.
The magnet field is provided by the hybrid magnetic system, which is toroidal
in the barrel part and solenoidal in the endcaps. The detector is
expected to have an average momentum resolution of about 1--2\% and
the designed energy resolution of ECAL is $5\%/\sqrt{E}$. The experiment
is expected to accumulate about 20 million $J/\psi$ events
during one year of smooth data taking.

%Compared to the previous experiments SPD advantages are high statistics,
%close to $4\pi$ acceptance, good momentum resolution, open spectrometer.
%The designed performance should enable precise measurement of the $J/\psi$
%production cross-section, its $p_T$- and $x_F$-spectra,
%and, most importantly, $J/\psi$ polarization in the whole
%kinematic range.

\begin{figure}[h]
	\begin{center}
		\begin{minipage}{17pc}
			\begin{center}
				\includegraphics[width=\textwidth]{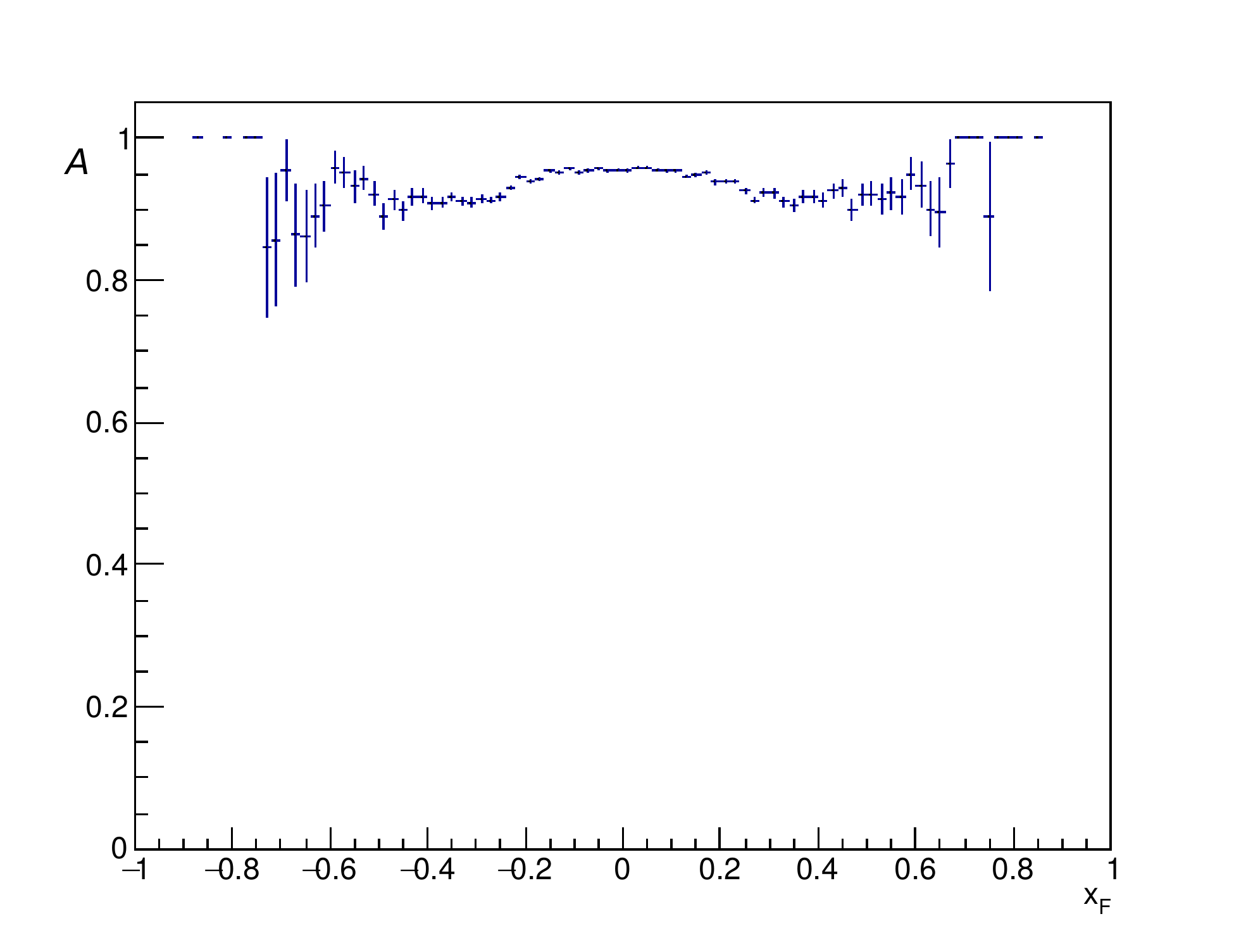}
				\caption{\label{fig:jpsi_xF} Detector acceptance for the $J/\psi$
					as a function of $x_F$.}
			\end{center}
		\end{minipage}
	    \hspace{1pc}
		\begin{minipage}{17pc}
			\begin{center}
				\includegraphics[width=\textwidth]{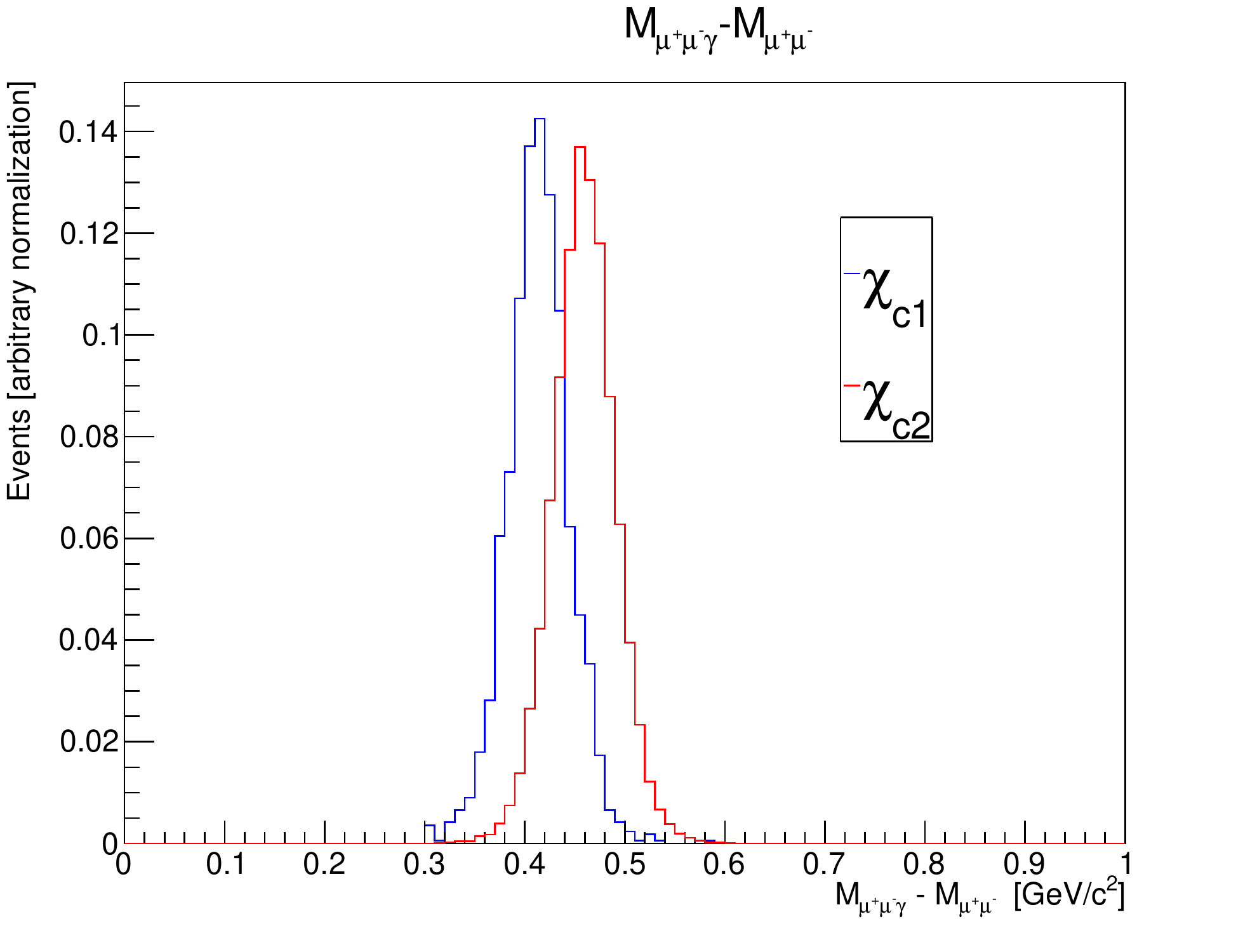}
				\caption{\label{fig:chi_dM}  $\Delta M = M_{\mu^+\mu^-\gamma} -
					     M_{\mu^+\mu^-}$ for $\chi_{c1}$ and $\chi_{c2}$.}
			\end{center}
		\end{minipage}
	\end{center}
\end{figure}

The basic event simulation takes into account the magnetic field and the material map
in the detector and is performed with the SPDRoot software. It shows that at the
collision energy of 26~GeV more than 90\% of muons produced in $J/\psi$ decays reach RS.
The detector acceptance for inclusive $J/\psi$ events is defined as a relative
fraction of events when both muons reaching RS.  It is shown in Fig.~\ref{fig:jpsi_pt}
and Fig.~\ref{fig:jpsi_xF} as a function of $p_T$ and $x_F$, respectively.
Keeping in mind the model fits from Fig.~\ref{fig:nrqcd-cem},
one can expect SPD to cover the $x_F$ region where
the contribution of the quark-antiquark annihilation becomes strong or dominant.

The $\chi_{c1}$ and $\chi_{c2}$ are reconstructed from the $\gamma J/\psi$
decay mode. The detector has good acceptance for the decay photons and
in about 80\% of such events muons reach RS and the photon reaches ECAL
and has energy above 100~MeV.
Preliminary generator-level studies show that ECAL energy resolution
may be sufficient to determine relative contributions of
$\chi_{c1}$ and $\chi_{c2}$. Fig.~\ref{fig:chi_dM} shows simulated
$\Delta M = M_{\mu^+\mu^-\gamma} -
M_{\mu^+\mu^-}$ distribution for $\chi_{c1}$ and $\chi_{c2}$.
Here charged track momentum and photon energy
smearing is introduced to simulate the detector response.
The feasibility of the unique measurement of $\chi_{cJ}$ decays
contribution to $J/\psi$ polarization is unclear and
will depend on the background.

\section{Summary}
Study of charmonia is a promising part of the SPD physics program.
The designed experiment will be capable to significantly improve
existing experimental knowledge of charmonia production by providing
precise and consistent measurement of its properties.
For unpolarized beams, it is of a special interest for the $\chi_{c1}$, $\chi_{c2}$, and $\psi(2S)$ states. This experimental input would be
crucial for the validation of
theoretical models (like CEM and NRQCD) and proper factorization
approaches at SPD energies. In case of polarized beams,
the high statistics and good detector performance will allow for a
precise measurement of spin asymmetries. In particular,
SPD will be able to validate the evidence for nonzero TSSA reported by the
PHENIX Collaboration. If a proper approach to charmonia production
is found and validated, it will provide new and significant information
on transverse structure of proton.

\section*{References}
\bibliography{jpsi}

\end{document}